\documentclass[prl,twocolumn,showpacs,floatfix,superscriptaddress]{revtex4}
\usepackage{amsmath,amsfonts}
\usepackage{graphicx}
\usepackage{color}
\usepackage{enumerate}

\newcommand{\beh}{\hat{\mathbf{e}}}
\newcommand{\bxh}{\hat{\mathbf{x}}}
\newcommand{\byh}{\hat{\mathbf{y}}}
\newcommand{\bzh}{\hat{\mathbf{z}}}
\newcommand{\bI}{\mathbf{I}}
\newcommand{\bJ}{\mathbf{J}}
\newcommand{\balpha}{\boldsymbol{\alpha}}
\newcommand{\bsigma}{\boldsymbol{\sigma}}
\newcommand{\up}{\uparrow}
\newcommand{\dw}{\downarrow}

\begin{document}


\title{Self-stabilizing temperature driven crossover between topological
and non-topological ordered phases in one-dimensional conductors}

\author{Bernd Braunecker}
\affiliation{SUPA, School of Physics and Astronomy,
University of St Andrews, North Haugh, St Andrews KY16 9SS, UK}
\author{Pascal Simon}
\affiliation{Laboratoire de Physique des Solides, CNRS UMR-8502,
             Universit\'{e} de Paris Sud, 91405 Orsay Cedex, France}

\date{\today}


\pacs{
74.20.Mn, 
71.10.Pm, 
75.30.Hx, 
75.75.-c  
}



\begin{abstract}
We present a self-consistent analysis of the topological superconductivity
arising from the interaction between self-ordered localized magnetic moments
and electrons in one-dimensional conductors in contact with a superconductor.
We show that due to a gain in entropy there exists a magnetically ordered
yet non-topological phase at finite temperatures that is relevant for
systems of magnetic adatom chains on a superconductor.
Spin-orbit interaction is taken into account, and we show that it causes a
modification of the magnetic order yet without affecting the topological
properties.
\end{abstract}


\maketitle


\emph{Introduction.}
Topological superconductors have received much attention recently,
partly because they host exotic low energy excitations such as Majorana bound states \cite{kane_RMP,alicea,leijnse},
whose non-Abelian statistics are attractive for topological quantum computation \cite{nayak,pachos:2012}.
As a remarkable feature, topological superconductivity can be created artificially by contacting specific
materials with a conventional $s$-wave superconductor. For instance, it arises at the interface between the
surface states of a three-dimensional topological insulator and a $s$-wave superconductor \cite{Fu-Kane};
 in one-dimensional (1D) semiconducting wires with a strong spin-orbit interaction (SOI) and
a Zeeman magnetic field with proximitized superconductivity \cite{lutchyn:2010,oreg:2010,mourik:2012,das:2012,deng:2012};
or in arrays of magnetic nanoparticles or magnetic adatoms on top of a superconducting surface
\cite{choy:2011,nadj-perge:2013,pientka:2013,pientka:2014,poyhonen:2014,rontynen:2014,kim:2014,heimes:2014,brydon:2015,ojanen:2015,schecter:2015b}
such as iron adatoms on lead \cite{nadj-perge:2014,pawlak:2015,ruby:2015}.

\begin{figure}
\centering
	\includegraphics[width=\columnwidth]{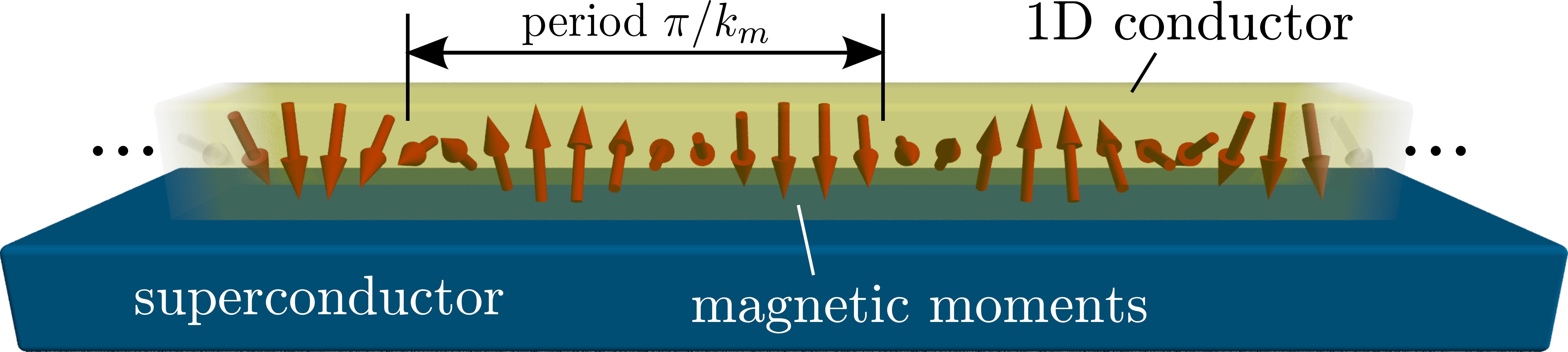}
	\caption{\label{fig:setup}
	Zoom on a 1D conductor with embedded magnetic moments on top of
	a superconductor. The magnetic moments self-order in the form
	a spiral order with spatial period $\pi / k_m$.}
\end{figure}

The systems we consider in this letter exhibit a topological phase emerging
from self-organization of magnetic moments embedded in 1D conductors with proximity induced superconductivity.
This situation may apply to semiconducting wires
with extrinsic magnetic impurities or intrinsic moments such as nuclear spins,
or a conducting wire made of magnetic adatoms on a superconducting surface.
Due to the Ruderman-Kittel-Kasuya-Yosida (RKKY) interaction mediated through the electrons,
the magnetic moments can undergo an ordering transition below a temperature $T^*$
and form a spiral with a spatial period characterized by the wave number $2k_m$
(see Fig.~\ref{fig:setup}) such that $k_m=k_F$, for $k_F$ the Fermi momentum.
This ordering mechanism was first demonstrated for normal
conductors \cite{braunecker:2009a,braunecker:2009b}, then conjectured
\cite{nadj-perge:2013} and self-consistently demonstrated
\cite{braunecker:2013,klinovaja:2013,vazifeh:2013} for the superconducting case.
These results were corroborated recently by showing
that the spiral order persists beyond the RKKY limit, and $k_m$ stays close
to $k_F$, as long as $k_F$ is away from commensurate band fillings
and the coupling strength $A$ between magnetic moments and electrons remains
smaller than the electron bandwidth \cite{schecter:2015,hu:2015}.

The locking of $k_m$ to $k_F$ has important consequences.
The magnetic spiral forms a periodic superstructure that
causes a part of the electrons to undergo a spin-selective Peierls
transition \cite{braunecker:2010} to a non-conducting spiral electron spin density wave,
whereas the remaining conducting electron states become helical (spin-filtered).
The induced superconductivity then becomes of the topological $p$-wave
type, and Majorana bound states appear at the two ends of the 1D wire.
A system with such a spiral magnetic order is indeed equivalent
\cite{braunecker:2010,choy:2011,kjaergaard:2012,martin:2012,jelena:2012,nadj-perge:2013,braunecker:2013,klinovaja:2013,vazifeh:2013,
pientka:2013,pientka:2014,poyhonen:2014,rontynen:2014,kim:2014,ojanen:2015,jelena:2015}
to the original proposals for topological superconductivity in nanowires \cite{lutchyn:2010,oreg:2010}.
Remarkably, by this mechanism, the topological superconducting phase emerges naturally
as the ground state without any fine tuning.
Although both the RKKY based and the nonperturbative approaches consistently predict
the locking condition $k_m \sim k_F$, it must be stressed that
the former results are based
on the further condition of large magnetic and superconducting gap energies,
whereas the latter apply only at zero temperature.

In this letter, we provide a general analysis
which incorporates entropy and thermal fluctuations in the non-zero temperature regime and we
show that there exists a previously unknown crossover to a
magnetically ordered yet non-topological phase.
Furthermore, we show that the spin-orbit interaction (SOI), which is genuinely
present in such systems either intrinsically or through interface effects, causes
a modification of the magnetic spiral but has no effect on the topological properties.

\begin{figure}
\centering
	\includegraphics[width=\columnwidth]{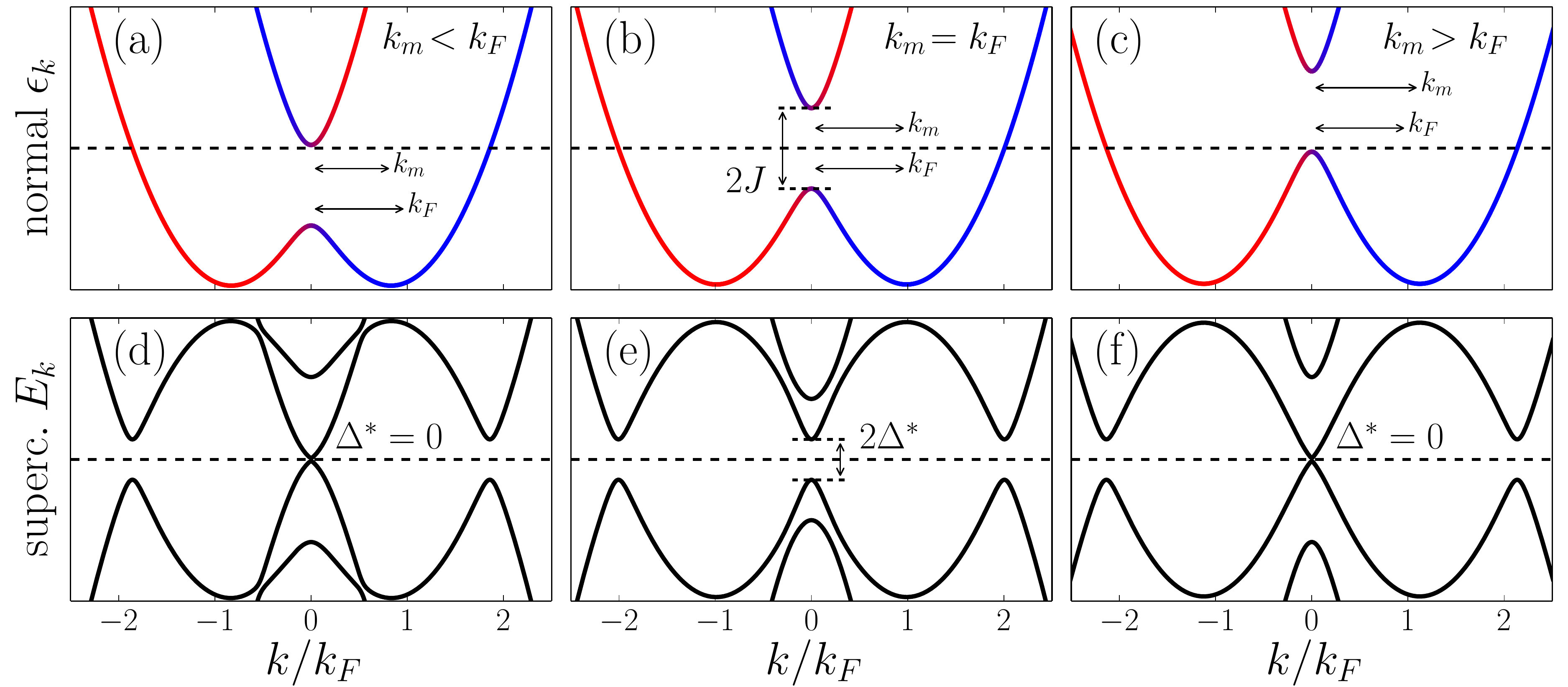}
	\caption{\label{fig:dispersions}
	Example of dispersion relations in the transformed basis obtained by the spin dependent
	shift of momenta $k \to k + \sigma k_m$ for which
	the spiral effective magnetic field becomes ferromagnetic and the Zeeman like
	gap $J$ opens at $k=0$ \cite{braunecker:2010,note_offset}.
	In the plots, the chemical potential $\mu$ remains constant, but $k_m$ varies.
	Top panels for the normal state
	($\epsilon_k$), with red and blue colors corresponding to opposite spin projections
	perpendicular to the spiral field.
	Bottom panels for the induced
	superconducting state ($E_k$, with $\Delta_s < J$).
	For $k_m = k_F$, the chemical potential $\mu$ (dashed line) lies in the middle
	of the $J$ gap (b) and the superconducting system is fully gapped (e). For smaller
	or larger $k_m$, the gap lies at lower or higher
	energies and $\mu$ eventually touches the upper (a) or lower gap edge (c).
	At both touching points the superconducting gap closes (d,f) and the state becomes non-topological.}
\end{figure}

\emph{ Heuristic considerations.}
The physical origin of the non-topological phase is illustrated in Fig. \ref{fig:dispersions}.
At low temperatures $T$, the thermodynamic ground state is determined by the
gain in electronic energy $E$ obtained by maintaining large magnetic ($J$)
and superconducting ($\Delta_s$) gaps, and the system adjusts $k_m$ to $k_F$.
As $T$ is raised, however,
the ground state is dictated by the free energy $F=E-TS$, and the entropy $S$
can play a decisive role. Indeed, if $k_m$ is lowered or raised to the values
as indicated in the left and right panels of Fig. \ref{fig:dispersions},
such that the chemical potential $\mu$ touches the bottom or top of a band,
the induced superconducting gap closes (for $J > \Delta_s$)
because the touched bands are fully spin polarized. The effective
dispersion arising from the superconducting case becomes gapless,
with a larger entropy than in the gapped case.
As a result, if $T$ is large enough, typically $k_B T \lesssim \Delta_s$ (with $k_B$
the Boltzmann constant),
the minimization of $F$ can be dominated by the enhancement of $S$, and the
thermodynamic ground state corresponds to situations (d) or (f) in Fig. \ref{fig:dispersions},
a topologically trivial yet magnetically ordered phase.

Not yet taken into account in this argument is the stability of this phase
upon thermal fluctuations of the magnetic moments.
As shown in \cite{braunecker:2009a,braunecker:2009b,braunecker:2013},
for both the ungapped and gapped cases
a mean-field description of the spin-wave fluctuations captures the correct
value of the ordering temperature and $T^* \propto |\chi_{2k_m}|$,
with $\chi_{2k_m}$ the transverse spin susceptibility at momentum $2k_m$.
Since $|\chi_{2k_m}|$ increases for a gapless dispersion, closing the
superconducting gap by moving $k_m$ away from $k_F$ causes furthermore
an enhanced stability against thermal fluctuations, provided
that the effect occurs at $T < T^*$. Obviously, the latter condition depends on the considered
material as specified below. For practical implementations, the condition
$k_B T^* \sim \Delta_s$ is {\it a priori} required for the topological self-tuning phase to
be accessible at high enough temperatures, and for $T$ close to $T^*$ the non-topological
ordered phase may indeed be favored. Yet we find that for semiconductor bands
with an effective mass as in Fig. \ref{fig:dispersions}, the value of $T^*$
remains generally still too low, such that $S$ would dominate $E$ only at
temperatures $T \gtrsim T^*$ where no order can persist anyway.

\begin{figure}
\centering
	\includegraphics[width=\columnwidth]{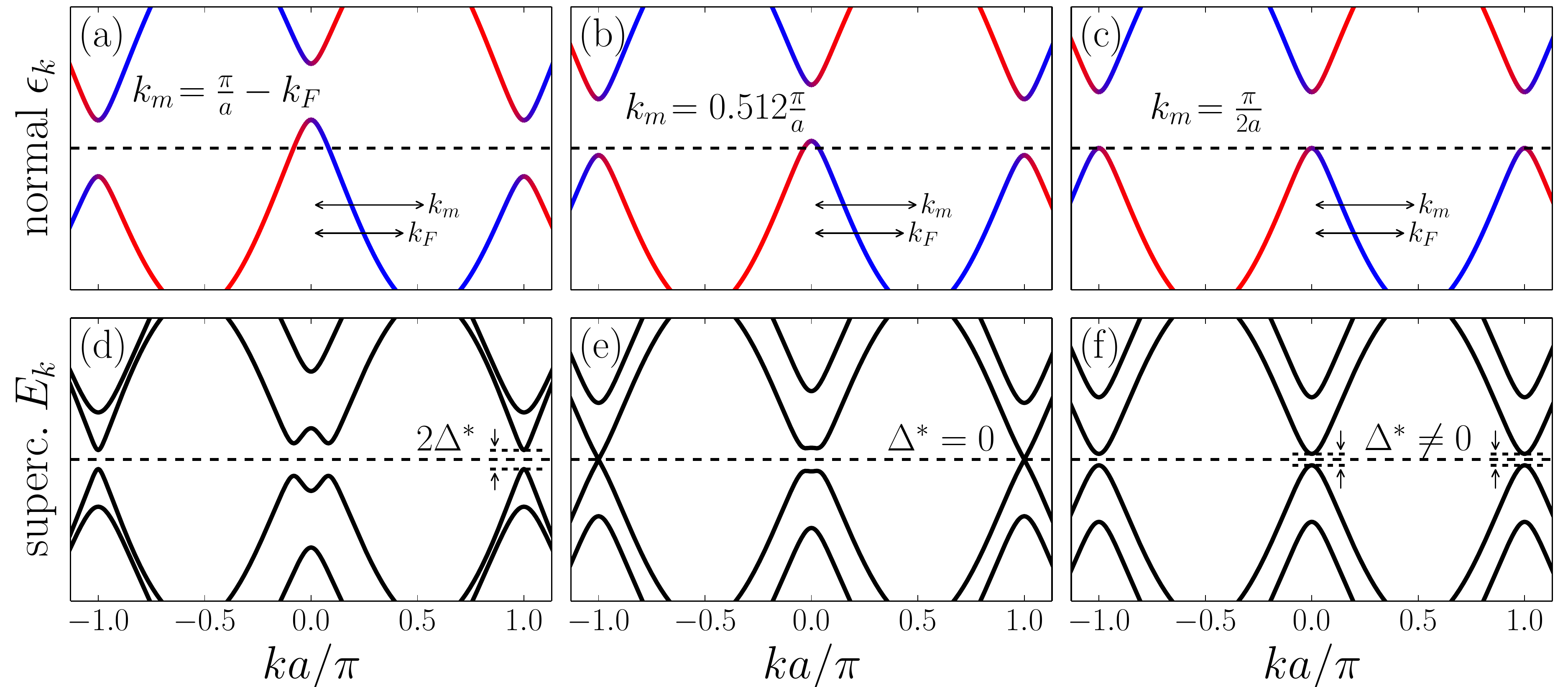}
	\caption{\label{fig:dispersions2}
	Dispersion relations as in Fig. \ref{fig:dispersions} for the tight binding
	model associated with Eq. (\ref{eq:H}) and corresponding to Fig. \ref{fig:F}.
	Parts (a,d) represent the
	low-temperature topological phase; parts (b,e) the crossover to the
	non-topological phase where the gap $\Delta^*$ closes; parts (c,f) the
	final high-temperature phase where $k_m = \pi/2a$
	and the system is magnetically ordered yet non-topological.
	}
\end{figure}

This situation changes drastically for tight binding systems such as
shown in Fig. \ref{fig:dispersions2}, which are the natural description for
adatom chains. Due to the cosine nature of the dispersion,
magnetic gaps appear at two points in the Brillouin zone, and $k_m$ can
self-adjust such that the superconducting-magnetic gaps $\Delta^*$ at both points
become equal and fulfill the condition $\Delta^* < \{k_BT^*, \Delta_s\}$ [see Fig. \ref{fig:dispersions2} (f)].
At $T \sim \Delta^*$ the effective doubling of thermally accessible states provides a doubling of
the value of $S$. This is sufficient to push the transition to the non-topological
to $T < T^*$ in precisely the systems that are most attractive for realizing
a self-sustained topological phase. Furthermore, the equality of the two gaps
$\Delta^*$ leads to $k_m = \pi/2a$ (with $a$ the lattice spacing),
which corresponds to an antiferromagnetic arrangement of the magnetic moments
if the latter are on the same lattice sites.

\emph{Quantitative analysis}.
For a quantitative investigation we consider
a quantum wire with induced superconductivity and embedded
magnetic moments, described by the Hamiltonian
\begin{eqnarray}
	&&\hspace*{-4mm}
	H
	= \sum_{k,\sigma}
	(\epsilon_{k} - \mu)
	c_{k,\sigma}^\dagger c_{k,\sigma}
	+
	\sum_{k,\sigma,\sigma'}
	(\balpha \cdot \bsigma)_{\sigma,\sigma'}
	k c_{k,\sigma}^\dagger c_{k,\sigma'}
	\label{eq:H}
\\
	&&\hspace*{-4mm}+
	\sum_k
	\bigl( \Delta_s c_{k,\up}^\dagger c_{-k,\dw}^\dagger + \text{h.c.} \bigr)
	+
	\sum_{k,q,\sigma,\sigma'} (\bJ_q \cdot \bsigma)_{\sigma,\sigma'} c_{k+q,\sigma}^\dagger c_{k,\sigma'}.\nonumber
\end{eqnarray}
Here $c_{k,\sigma}$ are the operators for electron with spin $\sigma =\up,\dw$ and
dispersion relation $\epsilon_k$.
$\mu$ is the chemical potential, $\Delta_s$ the induced superconducting gap,
$\bsigma = (\sigma_x,\sigma_y,\sigma_z)$ the vector of Pauli matrices,
and the vector $\balpha$ the effective SOI in the system,
arising from the sum of SOI contributions due to the internal structure of the wire or
to interface effects with the substrate.
The vectors $\bJ_q$ are the Fourier transforms of the chain of magnetic scatterers $\bJ_i = A \bI_i$ coupling to the electron
spin, where $\bI_i$ are the magnetic moments and $A$ is the coupling strength. The $\bJ_i$ are placed at positions
$r_i$ that can be irregular but are sufficiently dense with respect to $2\pi/k_F$ such that they can be considered as a continuum.

For $T=0$ and $\balpha = 0$, the ground state energy is minimized if the vectors $\bJ_i$ are confined
to an arbitrary two-dimensional plane, spanned by orthogonal unit vectors $(\beh_1,\beh_2)$,
in which they rotate as a spiral as a function of $r_i$,
$\bJ_i = J [ \cos(2 k_m r_i) \beh_1 + \sin(2 k_m r_i) \beh_2]$
\cite{choy:2011,kjaergaard:2012,martin:2012,nadj-perge:2013,braunecker:2013,klinovaja:2013,vazifeh:2013,kim:2014,schecter:2015,hu:2015}.
Choosing $\bxh, \byh$ such that $(\bxh,\byh) = (\beh_1,\beh_2)$,
the corresponding term in
the Hamiltonian becomes $\sum_k J ( c_{k+k_m,\up}^\dagger c_{k-k_m,\dw} + \mathrm{h.c.})$.
Letting
$c_{k+\sigma k_m,\sigma} \to \tilde{c}_{k,\sigma}$, $\epsilon_k \to \epsilon_{k\mp \sigma k_m,\sigma}$
\cite{braunecker:2010} produces a unitary transformation diagonalizing the Hamiltonian,
in which $J$ forms a uniform ferromagnetic coupling along the spin-$x$ direction. For a parabolic dispersion
$\epsilon_k=\hbar^2 k^2/2m$ with the band mass $m$ we obtain
Fig. \ref{fig:dispersions}.

Since the SOI term is linear in $k$, a $\balpha\neq 0$ produces a similar spin-dependent momentum
shift. For a quadratic dispersion
$\epsilon_k = \hbar^2 k^2/2m$ with $m$ the band mass,
and spin axes $\sigma_\alpha$ such that $\balpha\cdot\bsigma=\alpha \sigma_\alpha$,
the SOI can be absorbed by letting $\epsilon_k \to \epsilon_{k+\sigma_\alpha k_{SO}}$,
with $k_{SO} = \alpha m/\hbar^2$ \cite{braunecker:2010}.
If $\balpha$ is not perpendicular to the $(\beh_1,\beh_2)$ plane, the two shifts by $k_{SO}$ and
$k_m$ are not compatible, and the diagonalization of $H$ would generally mix all momenta.
Such modification of the long-ranged wave functions by the spiral order would cause an extensive
energy cost which is not favored energetically.
This can be avoided, however, by an alignment of the $\bJ_i$ spiral to the plane perpendicular to $\balpha$.
We thus define the spin directions such that $\balpha = \alpha \bzh$, and $(\beh_1,\beh_2) = (\bxh,\byh)$,
for which $k_{SO}$ and $k_m$ are parallel and can be directly added.
Remarkably, to maintain the optimal $k_m$ by minimizing the free energy, the spiral undergoes
an adjustment of $k_m$ to a $k_m'$ such that $k_m = k_m' - k_{SO}$. While the ``$-$'' sign arises from the choice
of spin axes, $k_m$ can have either sign, and so two spirals with opposite
helicities and different periods, $k_m' = \pm |k_m| + k_{SO}$, are possible.
Therefore even a large SOI has no further influence than the pinning of the plane of the magnetic
spiral together with the adjustment of $k_m'$, provided that $2\pi/k_m'$ does not
become smaller than the electron lattice spacing or the average spacing between the $\bJ_i$.
As long as $k_m=\pm k_F$ (up to $J$ dependent corrections that can be included),
a measurement of the period and plane of the magnetic spiral
could therefore give a direct measurement of ${\mathbf k}_{SO} \propto \balpha$.

Due to the extensive energy cost, there are furthermore no conical deformations out of the spiral plane \cite{kim:2014,reis:2014}.
In a spin-Nambu matrix representation
spanned by the vectors $(c^\dag_{k+k_m,\up}, c^\dag_{k-k_m,\dw}, c_{-k+k_m,\up}, c_{-k-k_m,\dw})$
the Hamiltonian takes then the form
\begin{equation} \label{eq:Htb}
	H =
	\sum_{k>0}
	\begin{pmatrix}
		\xi_{k-k_m} & J     & 0                    & \Delta_s     \\
		J           & \xi_{k+k_m}  & -\Delta_s     & 0            \\
		0           & - \Delta_s^* & -\xi_{-k-k_m} & -J           \\
		\Delta_s^*  & 0            & -J            & -\xi_{-k+k_m}
	\end{pmatrix}
	+ E_0,
\end{equation}
for $\xi_k = \epsilon_k - \mu$, $k_m = k_m'-k_{SO}$, $J_q = J \delta_{|q|,|2k_m'|}$,
and the restriction of the summation to $k>0$ to avoid state overcounting.
The energy offset
is
$
	E_0 = \sum_{k>0} \bigl[ \xi_{-k+k_m} + \xi_{-k-k_m} + 2 J \bigr],
$
and due to its $k_m$ dependence must be kept for comparison of different $k_m$.
The diagonalization of the matrix in Eq. \eqref{eq:Htb}, for $\xi_{-k} = \xi_k$,
leads to the energies $E_k^{\nu,\nu'} = \nu' E_{k,\nu}$, for $\nu,\nu' = \pm$,
with
$E_{k,\pm}^2 = J^2 + \Delta_s^2
+ \xi_{+,k}^2 + \xi_{-,k}^2
\pm
2 \sqrt{\Delta_s^2 J^2 + \xi_{+,k}^2 ( J^2 + \xi_{-,k}^2) }$
for
$\xi_{\pm,k} = (\xi_{k+k_m} \pm \xi_{k-k_m})/2$.
This leads to the ground state energy
\begin{equation}
	E = E_0 + \sum_{k>0,\nu,\nu'} E_{k}^{\nu,\nu'} f_k^{\nu,\nu'},
\end{equation}
and the entropy
\begin{equation}
	S = - k_B \sum_{k>0,\nu,\nu'}
	\Bigl[
		f_k^{\nu,\nu'}\ln\bigl(f_k^{\nu,\nu'}\bigr)
		+
		\bigl(1-f_k^{\nu,\nu'}\bigr) \ln\bigl(1-f_k^{\nu,\nu'}\bigr)
	\Bigr],
\end{equation}
for $f_k^{\nu,\nu'} = [1+\exp(E_k^{\nu,\nu'}/k_B T)]^{-1}$
the Fermi function. Notice that the sums are restricted to $k>0$ to avoid overcounting.
To proceed, we exclude the well-known scattering effects occurring at commensurate
filling factors that can cause orderings different from spirals \cite{schecter:2015,hu:2015}.

Minimizing the free energy $F=E-TS$ determines the ordering vector $k_m$.
However, analyzing only $F$ is incomplete to assure the stability of the
ordered phase since the long-wavelength spin-wave fluctuations smooth any magnetization at finite $T$
for a system of finite size \cite{braunecker:2009a,braunecker:2009b,braunecker:2013,klinovaja:2013,vazifeh:2013}.
Taking this condition into account, it was demonstrated \cite{braunecker:2009b,braunecker:2013}
that for any realistic system size the mean field result,
\begin{equation} \label{eq:T*}
	k_B T^* = 2 J^2 |\chi_{2k_m}| \sim \frac{J^2 a'}{\pi \hbar v_F} \ln\left(\frac{E_F}{\Gamma}\right),
\end{equation}
provides the ordering temperature $T^*$ for both the gapped and gapless cases,
where $\chi_{2k_m}$ is the static transverse spin susceptibility,
expressed in terms of the Fermi energy $E_F = \hbar v_F k_F/2$, the Fermi velocity $v_F$, and
the short distance cutoff $a'$ (limited by lattice spacing $a$ or distance between the $\bJ_i$).
If $\Delta^*$ is the gap as indicated in Figs. \ref{fig:dispersions} and \ref{fig:dispersions2}
the energy $\Gamma$ is roughly set by $\max(\Delta^*, k_B T)$ and its value reflects the
transition between the gapped ($\Delta^* > k_B T$) and gapless regimes ($\Delta^* < k_B T$).
As a consequence, if $k_m$ departs from $k_F$ and $\Delta^*$ shrinks, the value of $T^*$
initially grows but then saturates at a self-consistent value where $\Gamma \sim k_B T^*$.
Electron interactions further modify the logarithm in Eq. \eqref{eq:T*} to a temperature dependent
power-law in the gapless case and can lead to a considerable further increase of $T^*$ \cite{braunecker:2009a,braunecker:2009b}.
We note that Eq. \eqref{eq:T*} results from analyzing fluctuations arising from
the RKKY interaction. While the RKKY limit is insufficient to characterize the ground state
and $F$ needs to be used, it correctly encodes the fluctuations away from the ground state
configuration for $J < a'/\hbar v_F$.

\emph{Application to various systems.}
Since the transition to the non-topological phase occurs at $k_B T \lesssim \Delta_s$,
materials with large $k_BT^* \sim \Delta_s$ shall be considered.
With typical $\Delta_s \sim 0.1 - 1$ meV,
most systems with nuclear magnetic moments
have $k_BT^* \ll \Delta_s$,
hence a guaranteed topological phase, albeit a low transition temperature $T^*$.
For most 1D wire situations (including InAs \cite{braunecker:2013}), we indeed
find that $T^*$ remains too low to allow for a strong impact of the entropy, and the self-sustained
topological phases remain stable.

Dense chains of magnetic adatoms on a
superconducting substrate have a larger coupling constant $A$.
If neighboring adatom orbitals hybridize, the chains become conductors
of the type of Eq. \eqref{eq:H}. SOI effects are generally strong in such systems \cite{nadj-perge:2014,pawlak:2015,ruby:2015}.
However, this causes here just a mere rearrangement of the magnetic helix $k_m \to k_m'$.
While systems such as in \cite{nadj-perge:2014,pawlak:2015,ruby:2015} likely depend much on the direct
exchange interaction between neighboring moments, we focus here instead on the
case where RKKY dominates over the latter
(therefore our results do not a priori apply to \cite{nadj-perge:2014,pawlak:2015,ruby:2015}).

\begin{figure}
\centering
	\includegraphics[width=\columnwidth]{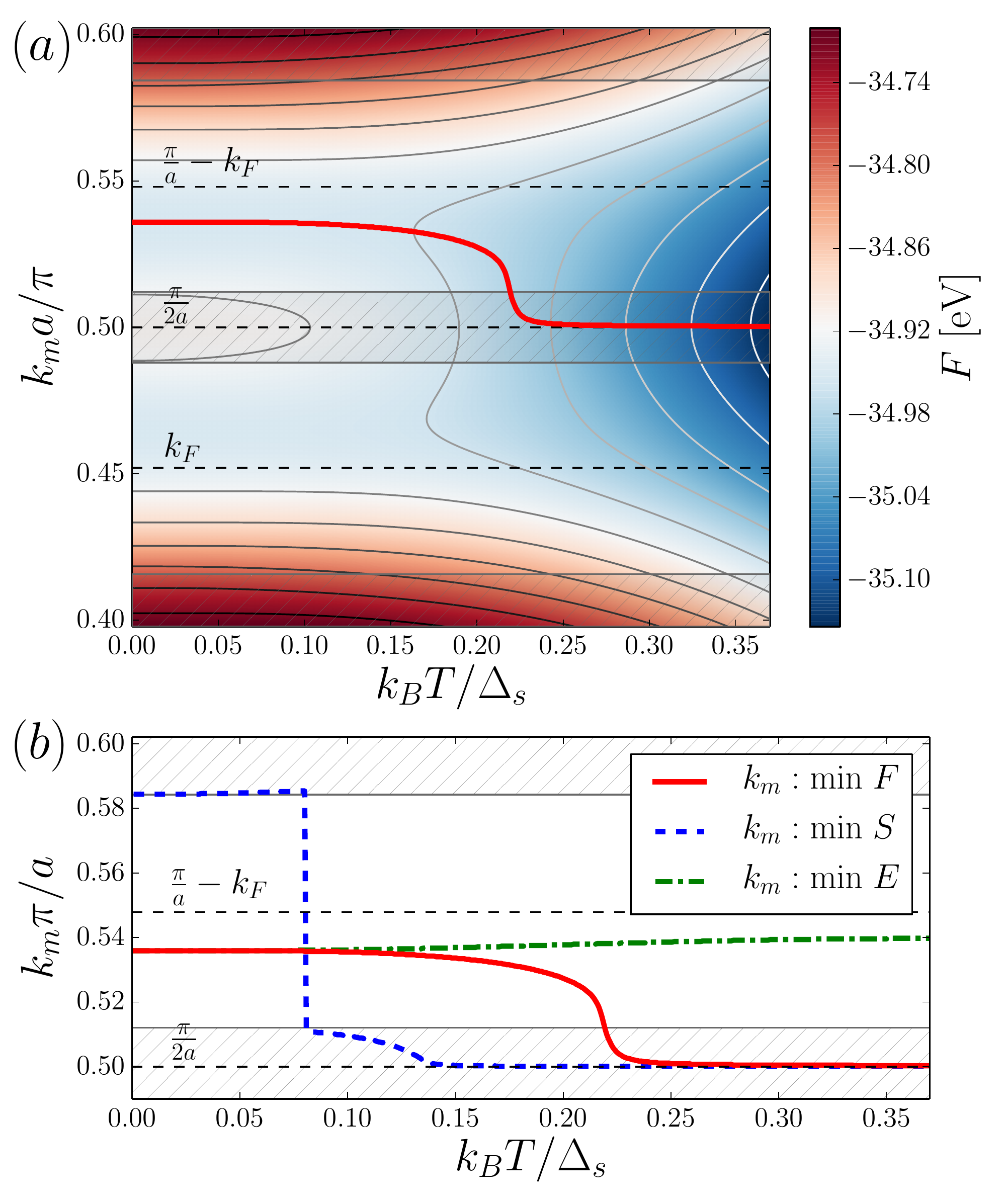}
	\caption{\label{fig:F}
	(a)
	Free energy $F$ for an adatom chain conductor of length $L =$ 1 $\mu$m
	as a function of total spiral momentum $k_m$
	(including SOI contributions) and temperature $T$ \cite{note_offset}.
	Contour lines complement the color coding.
	The minima $k_m(T)$ are marked by the
	red line and correspond to the ground state configuration.
	The values $k_F$, $\pi/a-k_F$, and $\pi/2a$ are marked by horizontal dashed lines.
	Parameters for the tight binding model are $t=10$ meV, $a=3$ \AA, $\Delta_s=2$ meV,
	$J=3$ meV, $L=1$ $\mu$m, $\mu = -0.3 t$ (7/20 filling).
	Stability of the order is ensured up to $k_B T^* = 0.37 \Delta_s$ (at the right plot limits).
	The gap $\Delta^*$ closes at the phase boundary between the topological (clear)
	and non-topological (hatched)
	regions. The system becomes non-topological
	at $k_B T \approx 0.22 \Delta_s$ and $k_m$ then stabilizes at $\pi/2a$, corresponding
	to an antiferromagnetic order.
	(b)
	Values of $k_m$ minimizing the free energy $F$ [same as in (a)], the entropy $S$, and the
	energy $E$. While $E$ alone favors a topological phase, the entropy $S$ favors
	a gapless phase, but is at higher temperatures further enhanced by tuning $k_m \to \pi/a$,
	and eventually dominates the minimum of $F$ at $k_B T > 0.22 \Delta_s$.
	For $k_m=\pi/2a$ the two identical gaps indicated in Fig. \ref{fig:dispersions2} (f) are
	$\Delta^* = 0.30 \Delta_s$.
	}
\end{figure}

As demonstrated in \cite{braunecker:2013}, with $A \sim 0.5$ meV \cite{menzel:2012}, we obtain
$k_B T^* > \Delta_s \sim 1$ meV for the gapped phase.
Requirement for such large $T^*$ is, as seen
in Eq. \eqref{eq:T*} a large prefactor $a'/\hbar v_F$, which means a rather small bandwidth
or a renormalization of the Fermi velocity \cite{Pientka:2015}.
A tight binding model is therefore suitable, in which the
factor $\hbar v_F /a'$ is replaced by the hopping integral $t$,
and the dispersion relation is $\epsilon_k = - 2 t \cos(a k)$.
Consequently, a minimum of $F$ at a shift $k_m$
has a particle-hole reversed minimum corresponding to the shift $(\pi/a-k_m)$. Away from half
filling, $k_F \neq \pi/2a$, both minima are inequivalent, yet for $k_F$ not too far from $\pi/2a$ they
can lie energetically close enough together such that further entropy can be gained by tuning $k_m$ through the
topological boundary and pin it to $k_m = \pi/2a$, an antiferromagnetic order at which
the system is non-topological and has small gaps but with both minima contributing to $S$.
In combination with a small bandwidth, a large $\Delta_s$, and $J \gtrsim \Delta_s$, the
entropy can become large enough to dominate $F$. An example is given in
Fig. \ref{fig:F}, showing $F$ as a function of $k_m$ and $T$. The values $k_m(T)$ minimizing $F(T)$ are
indicated by the red line. At low temperatures $k_m$ lies near, but not on $(\pi/a-k_F)$ since $J$ and $\Delta_s$
cause a significant band deformation due to the small band width.
At $k_B T \approx 0.22 \Delta_s$ we observe a crossing into the non-topological
region (indicated by the hatching), well below $k_B T^* \approx 0.37 \Delta_s$, and the
pinning to $k_m = \pi/2a$.

\emph{Conclusions}.
We have analyzed a 1D conductor with spin-orbit interaction coupled to a 1D chain of magnetic moments.
Through a self-consistent analysis taking into account
the full electronic free energy $F$ and the fluctuations about the ordered magnetic ground states,
we have determined the stability of the topological superconducting phase at finite temperature.
We showed that spin-orbit interaction causes only a pinning of the plane of the magnetic spiral and
an adjustment of its spatial period. Furthermore, in some situations especially met in
systems of magnetic adatoms, we demonstrated that there is a significant temperature range,
in which a magnetic order persist but the electronic state is non-topological.


\emph{Acknowledgments}.
We thank C. Carroll, J. Klinovaja, D. Loss and D. Morr for helpful discussions.
PS acknowledges support by the French Agence Nationale de la Recherche through the contract
ANR Mistral.



\begin{thebibliography}{99}
\bibitem{note_offset}
For a magnetic spiral, the shift $k \to k \mp \sigma k_m$ diagonalizes
the Hamiltonian without changing the energy, and hence the results for
different $k_m$ are directly comparable. The same type of transformation
also diagonalizes the SOI but adds an energy offset $\alpha^2 m/2$. However,
since the SOI is fixed for a given system, this offset is the same for
all further magnetic spirals and has no effect on the discussion.
\end{thebibliography}
\end{document}